 \documentclass[twocolumn,pra,showpacs]{revtex4}
 \draft
 \renewcommand{\a}{\alpha}
 \renewcommand{\b}{\beta}
 
 \newcommand{\cdj}{c^{\dag}_{j}}

 \newcommand{\cj}{c_{j}}

 \newcommand{\ad}{a^{\dag}}
 \newcommand{\p}{\prime}
 \renewcommand{\ap}{\alpha^{\p}}
 \renewcommand{\sp}{s^{\p}}
 \newcommand{\bp}{\beta^{\p}}
 \newcommand{\tp}{t^{\p}}
 
 \newcommand{\bd}{b^{\dag}}
 \newcommand{\cd}{c^{\dag}}
 \newcommand{\cs}{c^{s}}
 \newcommand{\hc}{\hat{c}}
 \newcommand{\asbt}{\alpha s,\beta t}
 \newcommand{\asbtp}{\alpha^{\p} s^{\p},\beta^{\p} t^{\p}}
 \usepackage{graphics}
          \begin{document}
           \title{A Representation of Complex Rational Numbers in Quantum mechanics}
          \author{Paul Benioff\\
           Physics Division, Argonne National Laboratory \\
           Argonne, IL 60439 \\
           e-mail: pbenioff@anl.gov}
           \date{\today}

          \begin{abstract}A representation of complex rational
          numbers in quantum mechanics is described that is not
          based on logical or physical qubits. It stems from
          noting that the $0s$ in a
          product qubit state do not contribute to the number.
          They serve only as place holders. The representation is
          based on the distribution of four types of systems on an
          integer lattice. The four types, labelled as positive real, negative
          real, positive imaginary, and negative imaginary, are
          represented by creation  and annihilation operators acting on
          the system vacuum state. Complex rational number
          states correspond to products of creation operators acting on the
          vacuum. Various operators, including those for the basic
          arithmetic operations, are described.  The representation used here is
          based on occupation number states and is given for bosons and
          fermions.
          \end{abstract}

          \maketitle

          \section{Introduction}

          Quantum computation and quantum information are subjects
          of much continuing interest and study. An initial impetus
          for this work was the realization that as computers became
          smaller, quantum effects would become more important.
          Additional interest arose with the discovery of
          problems \cite{Shor,Grover} that could be solved more
          efficiently on a quantum computer than on a classical
          machine. Also quantum information, and possibly quantum
          computation, \cite{Lloyd,Zizzi} is of recent
          interest in addressing problems related to cosmology and
          quantum gravity.

          In all of this work qubits (or qudits for d-dimesional
          systems) play a basic role. As quantum binary systems
          the states $|0\rangle,|1\rangle$ of a qubit represent the binary
          choices in quantum information theory.  They also
          represent the numbers $0$ and $1$ as numerical inputs
          to quantum computers. For $n$ qubits, corresponding
          product states, such as $|\underline{s}\rangle=
          \otimes_{j=1}^{n}|s(j)\rangle$ where $s(j)=0 \mbox { or
          }1,$ represent a specific $n$ qubit information state.
          Since they also represent numbers, \begin{equation}\label{qubitnum}
          |\underline{s}\rangle\rightarrow \sum_{j=1}^{n}\underline{s}(j)2^{j-1},
          \end{equation} they and their linear superpositions are inputs to
          quantum computers.

          It is clear that states of qubits are very important to quantum
          information theory. However, qubits and their states are not
          essential to the representation of numbers in quantum mechanics.
          This is based on the observation that in a state, such as
          $|10010\rangle$, the $0s$ do not contribute to the
          numerical value of the state. Instead they function more
          like place holders. What is important is the
          distribution of the $1s$ along a discrete lattice. This
          is shown by Eq. \ref{qubitnum} where the value of the
          number is determined by the distribution of $1s$ at the
          values of $j$ for which $\underline{s}(j)=1$.  The value, $0,$ of
          $\underline{s}$ at other locations contributes nothing.

          This suggests a different type of representation of
          numbers that does not use qubits. It is based instead on
          the distributions of $1s$ on an integer lattice. For
          example the rational  number $1001.01$ would be
          represented here as $1_{3}1_{0}1_{-2}.$ In quantum
          mechanics these states correspond to position eigenstates
          of a system on a discrete lattice or path where the positions
          are labelled by integers. Here the state $|j\rangle$ corresponds
          to the number $2^{j}$ and  $n$ product states, such as $|j_{1},\cdots
          j_{n}\rangle,$ correspond to $\sum_{k=1}^{n}2^{j_{k}}.$

          Here the representation of rational numbers
          corresponds to those represented by finite strings
          of binary digits or qubits and not as pairs of such
          strings. This representation is much easier to
          use and corresponds to that  used in computers.  It also
          is dense in the set of all rational
          numbers. For quantum states this means that any
          representation of all nonnegative rational numbers as quantum states
          would be approximated arbitrarily closely by a finite qubit
          $|\underline{s}\rangle$ states or states of the form
          $\otimes_{j\epsilon 1_{s}}|j\rangle$ where $1_{s} =
          \{j|\underline{s}(j)=1\}.$ In what follows this type
          of state will be referred to as a rational number state.

          This representation is sufficient to describe
          nonnegative rational numbers in quantum
          mechanics. There are several ways to extend the
          treatment to include negative and imaginary
          rational  numbers. These range from one type
          of system with two internal binary degrees of freedom
          to four different types of systems.  Here an intermediate
          approach is taken in which two types of systems which
          have an internal binary degree of freedom are
          considered. The two internal degrees of freedom
          correspond to positive and negative and the two types of
          systems correspond to real and imaginary. An example of
          such a number in the representation considered here is
          $r_{+,5}\: r_{-,3}\: i_{-,-2}\: i_{+,4}$

          The goal of this paper is to use these ideas to give a
          quantum mechanical representation of complex rational
          numbers. Since states with varying numbers of $r+,r-,i+,i-$
          systems will be encountered, a Fock  space representation
          is used. Both bosons and fermions will be considered.

          The emphasis of this work is to describe a set of
          quantum states that can be shown to represent complex
          rational numbers.  This requires definitions of the
          basic arithmetical operations used in the axiomatic
          definitions of rational numbers and showing that the
          states have the desired properties.

          An additional emphasis is that the state descriptions
          and properties must be relatively independent of the
          complex rational numbers that are part of
          the complex number field $C$ on which the
          Fock space is based. This means that the description
          will not be based on a map from quantum states to $C$
          that is used to define arithmetic properties of the
          states. Instead the states and their properties will be
          described independent of any such map.

          It will be seen that the representation used here is more compact with
          simpler representations of the basic arithmetic operations
          than those based on qubit states  with "binal" points, e.g.,
          of the form $|\pm 10010.011\rangle$ \cite{BenRNQM,BenRNQMALG}.
          It also extends the representation to complex numbers
          which was not done in the earlier work.
          Another (slight) advantage is that linear superpositions
          of states containing just one system are not entangled
          in the representation used here.  This is not the case
          for the qubit representation with $0s$ present.  An
          example is the Bell state $(1/\sqrt{2})(|10\rangle\pm
          |01\rangle).$ Here this state is $(1/\sqrt{2})(|1\rangle\pm|0\rangle),$
          which is not entangled. In this state $0$ and $1$ are the locations
          of the $1s$ in the qubit state.  This advantage is lost
          when one considers states with more than one system present.

           Another advantage of the representation
           described here is that it may suggest new physical
           models for quantum computation that are not qubit
           based.  Whether this is the case or not must await
           future work.

          The use of Fock spaces to describe quantum computation
          and quantum information is not new.  It has been used to
          describe fermionic \cite{Kitaev,Ozhigov} and parafermionic
          \cite{Lidar} quantum computation,
          and quantum logic \cite{Gudder,DChiara}.  The novelty of the
          approach taken here is based on a description of complex
          rational string numbers that is not based on logical or
          physical qubits. In this sense if differs from
          \cite{Kassman}. It also emphasizes basic arithmetic
          operations instead of quantum logic gates.  Also both
          standard and nonstandard  representations of numbers are
          described. These follow naturally from the occupation
          number description of quantum states.

          Details of the description of the complex rational states
          are given in the next  three sections. The a-c operators are described
          in Section \ref{CRSS}. The next section gives
          properties of these and other operators and their use to
          describe complex rational
          states. Section \ref{BAORSS} describes the
          arithmetic operations of addition, multiplication, and
          division to any finite accuracy. The last section
          summarizes some advantages of the approach used here.
          Also a possible physical model of standard and
          nonstandard numbers as pools of four types of Bose
          Einstein condensates along an integer lattice is briefly
          discussed.

          \section{Complex Rational States}\label{CRSS}

          The representation of complex rational  states used here
          is based on the notion of creating  and annihilating
          two types of systems, at various
          locations. One type is used for real rational states
          and the other for imaginary states. For bosons the degrees of
          freedom associated with each type consist of a binary
          internal degree, denoted by $+,-$, and a location $j$ on an
          integer labelled lattice.

          The creation operators for bosons are
          $\ad_{+,j},\ad_{-,j}, \bd_{+,j},\bd_{-,j}$.  The $a$  operators
          create and annihilate bosons in states corresponding respectively
          to positive and negative real rational  states.  The $b$
          operators play the same role for imaginary states.
          The state $|0\rangle$ is the vacuum state.

          In this representation,  the states $\ad_{+,j}|0\rangle,
          \ad_{-,j}|0\rangle$ show an a (real) boson in states
          $+,-$ at site $j$. The states $\bd_{+,j}|0\rangle,\bd_{-,j}
          |0\rangle$  show a b (imaginary) boson in states $+,-$
          at site $j$. In the order presented these states correspond to
          the numbers $2^{j},-2^{j},i2^{j},$ and $-i2^{j}.$

          For fermions the creation and annihilation operators
          have an additional variable $h =0,1,2,\cdots.$  This
          extra variable is needed to make fermions with the
          same sign and $j$ value distinguishable. Thus boson
          states of the form $\ad_{+,j}\ad_{+,j}|0\rangle$
          become $\ad_{+,h,j}\ad_{+,h^{\p},j}|0\rangle$ where $h\neq
          h^{\p}.$  Note that, as far as number properties are
          concerned, $h$ is a dummy variable in that
          $\ad_{+,h,j}|0\rangle$ and $\ad_{+,h^{\p},j}|0\rangle$
          both represent the same number. However in any physical
          model it would represent some physical property.

          One can also form  linear superpositions of these
          states. Simple boson examples and their equivalences in the
          usual qubit based binary notation are,
           \begin{equation}\label{a-ccorr}
          \begin{array}{l}(1/\sqrt{2})(\ad_{+,j}\pm \ad_{-,j})|0\rangle
          =(1/\sqrt{2})(|1\underline{0}^{j}\rangle\pm|-1\underline{0}^{j}\rangle)
          \\ (1/\sqrt{2})(\ad_{+,j}\pm \ad_{-,k})|0\rangle =(1/\sqrt{2})(|1\underline{0}^{j}
          \rangle\pm|-1\underline{0}^{k}\rangle) \\ (1/\sqrt{2})(\ad_{+,k} \pm\bd_{+,j}) =
          (1/\sqrt{2})(|1\underline{0}^{k}
          \rangle\pm|i1\underline{0}^{k}\rangle) \\ (1/\sqrt{2})(1\pm \bd_{-,j})|0\rangle
          =(1/\sqrt{2})(|0\rangle\pm|-i1\underline{0}^{j}\rangle)
          \\ (1/\sqrt{2})(1\pm \ad_{+,j})|0\rangle =(1/\sqrt{2})(|0\rangle\pm
          |1\underline{0}^{j}\rangle).\end{array}\end{equation}
          Here $\underline{0}^{j}$ represents a string of $j$ $0s.$

          These states show one advantage of the representation used here
          in that those on the left are valid for any value of $j$ or
          $k$. The usual binary representations on the right are
          valid only for $j,k\geq 0.$ Note also that $-$ and $i$
          inside the qubit states denote the type and sign of the
          number.  They are not phase factors multiplying the
          states.

          \section{ Occupation Number States}\label{RNSSPACO}

          The first step in representing states as products of
          creation operators acting on $|0\rangle$ is to give the commutation
          relations. Let $\cdj,\hat{c}^{\dag}_{j},c_{j},$
          and $\hat{c}_{j}$ be variable a-c operators where
          $\cd$ and $\hat{c}^{\dag}$ can take any one of
         the four values $\ad_{+},\ad_{-},\bd_{+},\bd_{-}.$ Using
         these the boson commutation relations can be given as \begin{equation}
          \label{ccomm} [c_{j} ,\cd_{k} ]=\delta_{j,k}
          \hspace{1cm} [\cdj ,\hat{c}^{\dag}_{k}] = [c_{j}
         ,\hat{c}_{k}] =0.\end{equation} The first
         equation stands for four equations as $\cd$ has any one
         of four values. Each of the next two equations stands for
         $16$ equations as $\cd$ and $\hat{c}^{\dag}$ each have
         any one of four values.

           For fermions  the anticommutation relations are given
           by\begin{equation}\label{canticomm} \{c_{g,j} ,\cd_{h,k}
          \}=\delta_{j,k}\delta_{g,h} \hspace{0.5cm}
          \{ \cd_{g,j} ,\hc^{\dag}_{h,k} \}  = \{c_{g,j}
         ,\hc_{h,k}\} =0 \end{equation} where
         $\{c,d\}=cd+dc.$ There are two sets of these relations,
         one for $\cd$ and $\hc^{\dag}$ each having the values
         $\ad_{+},\ad_{-}$ and the other for $\cd$ and
         $\hc^{\dag}$  with the values $\bd_{+}$ or $\bd_{-}.$  Note that
         because the $a$ and $b$ systems are two different types of
         fermions, \emph{commutation} relations hold between their
         operators, as in $[\ad_{+,g,j},\bd_{+,h,k}] =0,$ etc..

         A complete basis set of states can be defined in terms of
         occupation numbers of the various boson or fermion states. A general
         basis can be defined as follows:  Let $n_{r},m_{r},n_{i},m_{i}$
         be any four functions that map the set of all integers to the
         nonnegative integers. Each function has the value $0$
         except possibly on finite sets of integers. Let
          $s,s^{\p},t,t^{\p}$ be the four finite sets of integers
          which are the nonzero domains, respectively, of the four
          functions. Thus $n_{r,j}\neq 0[=0]$ if $j\epsilon s[j
          \mbox{ not in }s],$  $m_{r,j}\neq 0[=0]$ if $j\epsilon s^{\p}
          [j \mbox{ not in }s^{\p}],$ etc..

         Let $\bigcup s,t$ be the set of all integers in one or more of the four sets.
         Then a general boson occupation number state has the form
         \begin{equation}\label{occno}
         |n_{r},m_{r},n_{i},m_{i}\rangle = \prod_{j\epsilon \cup
         s,t}|n_{r,j},m_{r,j}n_{i,j}m_{i,j}\rangle\end{equation}
         where $|n_{r,j},m_{r,j}n_{i,j}m_{i,j}\rangle$ the
         occupation number state for site $j$ is given by
         \begin{equation}\label{occnost}\begin{array}{l}
         |n_{r,j},m_{r,j}n_{i,j}m_{i,j}\rangle=\frac{1}{N(n,m,r,i,j)} \\
         \hspace{1cm}\times (\ad_{+,j})^{n_{r,j}}
         (\ad_{-,j})^{m_{r,j}}(\bd_{+,j})^{n_{i,j}}(\bd_{-,j})^{m_{i,j}}|0\rangle.
         \end{array}\end{equation}  The normalization factor
         $N(n,m,r,i,j)=(n_{r,j}!m_{r,j}!n_{i,j}!m_{i,j}!)^{1/2}.$
         Note that the product $\prod_{j\epsilon \cup
         s,t}$ denotes a product of creation operators, and not
         a product of states.

         The interpretation of these states is that they are the
         boson equivalent of \emph{nonstandard} representations of
         complex rational numbers as distinct from \emph{standard}
         representations. (This use of standard and nonstandard is
         completely different from standard and nonstandard numbers
         described in mathematical logic \cite{Chang}.) Such nonstandard
         states occur often
         in arithmetic operations and will be encountered later on.
         They correspond to columns of binary numbers where
         each number in the
         column is any one of the four types, positive real,
         negative real, positive imaginary, and negative
         imaginary. In a boson representation individual systems,
         are not distinguishable. The only measurable properties
         are the number of systems of each type  $+1,-1,+i,-i$ in
         the single digit column at  each site $j.$

         An example would be a computation in which one
         computes the value of the integral $\int_{a}^{b}f(x)dx$
         of a complex valued function  $f$ by computing in parallel,
         or by a quantum computation, values of $f(x_{h})$ for
         $h=1,2,\cdots,m$ and then combining the $m$ results to
         get the final answer. The table, or matrix, of $m$ results
         before combination is represented here by a state
         $|n_{r},m_{r},n_{i},m_{i}\rangle$ where
         $n_{r,j},m_{r,j}n_{i,j},m_{i,j}$ give the number of
         $+1's$, $-1's,$ $+i's$, and $-i's$ in the column at site $j.$ This is
         a nonstandard representation because it is numerically
         equal to the final result which is a standard
         representation consisting of one real and one imaginary rational
         string number, often represented as a pair, $u,iv$.

         The equivalent fermionic representation for the
         state $|n_{r},m_{r},n_{i},m_{i}\rangle$ is based
         on a fixed ordering of the a-c operators.
         In this case the product $(\ad_{+,j})^{n_{r,j}}$ becomes
         $\ad_{+1,j}\cdots\ad_{+h,j}\cdots\ad_{+n_{r,j},j}$
         with similar replacements for $(\ad_{-,j})^{m_{r,j}},
         (\bd_{+,j})^{n_{i,j}},(\bd_{-,j})^{m_{i,j}}.$
         Each component state
         $|n_{r,j},m_{r,j},n_{i,j},m_{i,j}\rangle$
         in Eq. \ref{occnost} is given by
         \begin{equation}\label{occferm}\begin{array}{c}
         |n_{r,j},m_{r,j},n_{i,j},m_{i,j}\rangle=
         \ad_{+n_{r,j},j}\cdots \ad_{+1,j}\ad_{-m_{r,j},j}\cdots \\
         \ad_{-1,j}  \bd_{+n_{i,j},j}\cdots\bd_{+1,j}
         \bd_{-m_{i,j},j}\cdots\bd_{-1,j}|0\rangle\end{array}\end{equation}
         The final state is given by an ordered product over the $j$ value,
         \begin{equation}\label{occnoferm}|n_{r},m_{r},n_{i},m_{i}\rangle =
         \prod_{j\epsilon \cup s,t}J|n_{r,j},m_{r,j}n_{i,j}m_{i,j}
         \rangle.\end{equation} Here $J$ denotes a $j$ ordered
         product where factors with larger values of $j$ are to
         the right of factors with smaller $j$ values. The choice
         of ordering, such as that used here in which
         the ordering of the $j$ values is the opposite of that
         for the $h$ values which increase to the left as in Eq.
         \ref{occferm}, is arbitrary.  However, it must remain fixed
         throughout.

\begin{figure}[t]\begin{center}\vspace{1cm}
           \resizebox{100pt}{100pt}{\includegraphics[230pt,120pt]
           [530pt,420pt]{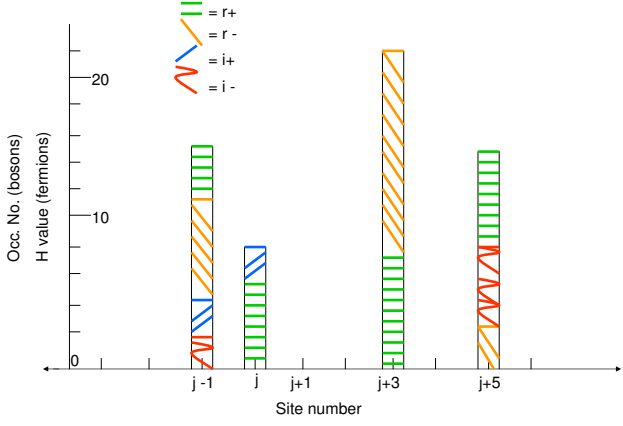}}\end{center}
           \caption{Example of a nonstandard complex rational
           state for bosons and fermions for $4$ occupied
           $j$ values.  At each site $j$
           the vertical bar shows the occupation numbers (bosons)
           or extent of $h$ values (fermions) with different colors
           and line slopes showing each of the four types of systems.
           For instance the bar at site $j-1$ shows  $13\; r+,$
           $11\; r -,$ $4\; i+,$ and $2\; i -$ systems and the bar at
           site $j$ shows $5\; r+$ and $8\; i+$ bosons. For fermions
           the ordinate labels the ranges of the $h$ values for each
           type.}\label{fig1} \end{figure}

         An example of a nonstandard representation is illustrated in
         Figure \ref{fig1} for both bosons and fermions. The
         integer values of $j$ are shown on the abcissa. The
         ordinate shows the  boson occupation numbers for each type of
         system. Fermions are represented  as two types
         of systems each with two internal states $(+,-)$ on a two
         dimensional lattice with $j$ any integer and $h$ any nonnegative
         integer. The ordinate shows the range of $h$
         values from $0$ to $n_{r,j},m_{r,j},n_{i,j},m_{i,j}$ for
         each of the four types.

         The above shows the importance of nonstandard
         representations, especially in cases where a large amount
         of data or numbers is generated which must be combined
         into a single numerically equivalent complex rational
         number. This requires  definition of standard
         complex rational number states and of properties
         to be satisfied by any conversion process.

         For bosons a standard complex rational
         state has the form of Eq. \ref{occnost} where
         one of the functions $n_{r},m_{r}$ and one of $n_{i},m_{i}$
         has the constant value $1$ on their nonzero domains. The other two
         functions are $0$. The four possibilities are \begin
         {equation}\label{stdbos}\begin{array}{c}
         |\underline{1}_{s},0,\underline{1}_{t},0\rangle =(\ad_{+})^{s}
         (\bd_{+})^{t}|0\rangle \\|\underline{1}_{s},0,0,
         \underline{1}_{t^{\p}}\rangle =(\ad_{+})^{s}
         (\bd_{-})^{t^{\p}}|0\rangle\\|0,\underline{1}_{s^{\p}},
         \underline{1}_{t},0\rangle = (\ad_{-})^{s^{\p}}(\bd_{+})^{t}|0\rangle
         \\|0,\underline{1}_{s^{\p}},0,\underline{1}_{t^{\p}}\rangle =
         (\ad_{-})^{s^{\p}}(\bd_{-})^{t^{\p}}|0\rangle.
         \end{array}\end{equation} Here $(\ad_{+})^{s}=\prod_{j\epsilon
         s}\ad_{+,j}$ and $\underline{1}_{s}$ denotes the constant
         $1$ function on $s$, etc. Pure real or imaginary  standard
         rational states are included if $t,\;t^{\p}$ or
         $s,\;s^{\p}$ are empty.  If $s,\;s^{\p},\; t,\; t^{\p}$ are
         all empty one has the vacuum state $|0\rangle.$  Note that
         Eq. \ref{stdbos} also is valid for fermions  with the
         replacements\begin{equation}\label{stdfer}\begin{array}{l}
        (\ad_{\a} )^{s}\rightarrow \ad_{\a,1,j_{1}}\ad_{\a,1,j_{2}}
        \cdots\ad_{\a,1,j_{|s|}} \\ (\bd_{\b})^{t}\rightarrow
        \bd_{\b,1,k_{1}}\bd_{\b,1,k_{2}}\cdots\bd_{\b,1,k_{|t|}}.\end{array}
        \end{equation} Here $\a=+,-$, $\b=+,-$, and
        $s=\{j_{1},j_{2},\cdots,j_{|s|}\},\;
        t=\{k_{1},k_{2},\cdots,k_{|t|}\}$. Also
        $j_{1}<j_{2}<\cdots<j_{|s|},\;k_{1}<k_{2}<\cdots<k_{|t|},$
        and $|s|,|t|$ denote the number of integers in $s,t.$

         Standard states are quite important.  All theoretical
         predictions as computational outputs, and numerical
         experimental results are represented by standard real
         rational states. Nonstandard representations occur
         during the computation process and in any situation where
         a large amount of numbers is to be combined. Also qubit states
         correspond to standard representations only.

         This shows that it is important to describe the numerical
         relations between nonstandard representations and standard
         representations and to define numerical equality between states.
         To this end let \begin{equation}\label{Nequ}
         |n_{r},m_{r},n_{i},m_{i}\rangle
         =_{N}|n^{\p}_{r},m^{\p}_{r},n^{\p}_{i},m^{\p}_{i}\rangle\end{equation}
         be the statement of $N$ equality between the two
         indicated states. This statement is satisfied if two
         basic equivalences are satisfied.
         For bosons the two $N$ equivalences are
         \begin{equation}\label{abdjabj} \ad_{+,j}\ad_{-,j}=_{N}\tilde{1};\;\;\;\;
         \bd_{+,j}\bd_{-,j}=_{N}\tilde{1}\end{equation}
         and \begin{equation}\label{abjabj}\begin{array}{l} \ad_{\a,j}\ad_{\a,j}=_{N}
         \ad_{\a,j+1}\;\;\;\; a_{\a,j}a_{\a,j}=_{N}a_{\a,j+1} \\ \bd_{\b,j}\bd_{\b,j}=_{N}
         \bd_{\b,j+1}\;\;\;\; b_{\b,j}b_{\b,j}=_{N}b_{\b,j+1}.\end{array}\end{equation}

         The first pair of equations says that any  state that
         has one or more $+$ and $-$ systems of either the $r$ or $i$ type at a site $j$ is
         numerically equivalent to the state with one less $+$ and
         $-$ system  at the site $j$ of either type. This is the
         expression here of $2^{j}-2^{j}=i2^{j}-i2^{j}=0.$
         The second set of two pairs, Eq. \ref{abjabj}, says that any
         state with two systems of the same type and in the same
         internal state at site $j$, is numerically equivalent to a
         state without these systems but
         with one system of the same type and internal state at
         site $j+1.$ This corresponds to $2^{j}+2^{j}=2^{j+1}$ or
         $i2^{j}+i2^{j}=i2^{j+1}.$

         From these relations one sees that any process whose
         iteration preserves $N$ equality according to Eqs.
         \ref{abdjabj} and \ref{abjabj} can be used to
         determine if Eq. \ref{Nequ} is valid for two different
         states. For example if \begin{equation}\label{jj1}
         |n_{r},m_{r},n_{i},m_{i}\rangle
         =a_{+,j}a_{-,j}|n^{\p}_{r},m^{\p}_{r},
         n^{\p}_{i},m^{\p}_{i}\rangle\end{equation} or \begin{equation}\label{jjj+1}
         |n_{r},m_{r},n_{i},m_{i}\rangle
         =\bd_{+,j+1}b_{+,j}b_{+,j}|n^{\p}_{r},m^{\p}_{r},
         n^{\p}_{i},m^{\p}_{i}\rangle,\end{equation} then Eq.
         \ref{Nequ} is satisfied.

         For fermions the corresponding $N$ equivalences are
         \begin{equation}\label{fabdjabj}
         \ad_{+,j,h}\ad_{-,j,h^{\p}}=_{N}\tilde{1};\;\;\;\;
         \bd_{+,j,h}\bd_{-,j,h^{\p}}=_{N}\tilde{1}\end{equation}
         and \begin{equation}\label{fabjabj}\begin{array}{l}
         \ad_{\a,h,j}\ad_{\a,h^{\p},j}=_{N}
         \ad_{\a,h^{\p\p},j+1}\;\;\;\; a_{\a,h,j}a_{\a,h^{\p},j}
         =_{N}a_{\a,h,^{\p\p},j+1} \\ \bd_{\b,h,j}\bd_{\b,h^{\p},j}=_{N}
         \bd_{\b,h^{\p\p},j+1}\;\;\;\; b_{\b,h,j}b_{\b,h^{\p},j}
         =_{N}b_{\b,h^{\p\p},j+1}.\end{array}\end{equation}   In Eq. \ref{fabjabj}
         $h\neq h^{\p}.$ Otherwise the values of $h,h^{\p},h^{\p\p}\geq
         1$ are arbitrary except that removal of fermions is
         restricted to occupied $h$ values and addition is
         restricted to unoccupied values. To avoid poking holes in
         the $h$ columns at each site $j,$ Fig. \ref{fig1},
         it is useful to restrict system removal to the maximum
         occupied h value  and system addition to the nearest
         unoccupied $h$ site. Numerically it does not matter where, in
         the $h$ direction, the fermions are added or removed.

         These equations have a meaning similar that that for the
         corresponding boson equations.  Eq. \ref{fabdjabj} says
         that any state given by Eqs. \ref{occferm} and
         \ref{occnoferm} is $N$ equal to a state with one
         $\ad_{+}$ and one $\ad_{-}$ fermion removed from site $j$
         i.e. $n_{r,j}\rightarrow n_{r,j}-1$ and
         $m_{r,j}\rightarrow m_{r,j}-1.$ A similar situation holds
         for removal of one $\bd_{+}$ and one $\bd_{-}$ fermion
         from site $j.$ Eq. \ref{fabjabj} says that a state with two
         $\ad_{+}$ or two $\ad_{-}$ fermions removed from site $j$
         is $N$ equal to a state with one $\ad_{+}$ or $\ad_{-}$
         fermion added to site $j+1.$ A similar situation holds
         for the $\bd_{+}$ or $\bd_{-}$ fermions.

         Corresponding to Eq. \ref{jj1}  one has
         the following: Let $|n_{r},m_{r},n_{i},m_{i}\rangle$ and
         $|n^{\p}_{r},m^{\p}_{r},n^{\p}_{i},m^{\p}_{i}\rangle$ be such that
         \begin{equation}\label{fjj1}
         |n_{r},m_{r},n_{i},m_{i}\rangle =R_{a,+,j}R_{a,-,j}|n^{\p}_{r}m^{\p}_{r}
         n^{\p}_{i}m^{\p}_{i}\rangle\end{equation} is satisfied
         where \begin{equation}\label{fRRjj1}\begin{array}{l}R_{a,+,j}=
         \sum_{h}a_{+,h+1,j}\ad_{+,h+1,j}a_{+,h,j}\\
         R_{a,-,j}=\sum_{h}a_{-,h+1,j}\ad_{-,h+1,j}a_{-,h,j},\end{array}
         \end{equation}then Eq. \ref{Nequ} is satisfied.  A
         similar statement holds if $b$ replaces $a$ in the above.
         The presence of the factors $a_{+,h+1,j}\ad_{+,h+1,j}$
         and $a_{-,h+1,j}\ad_{-,h+1,j}$ in Eq. \ref{fRRjj1} is to
         ensure that the $\ad_{+}$ and $\ad_{-}$ operators with
         the maximum $h$ values are deleted.

         Corresponding to Eq. \ref{jjj+1} one has that if $|n_{r},m_{r},n_{i},m_{i}\rangle$ and
         $|n^{\p}_{r},m^{\p}_{r},n^{\p}_{i},m^{\p}_{i}\rangle$
         satisfy \begin{equation}\label{fjjj+1}
         |n_{r},m_{r},n_{i},m_{i}\rangle
         =_{\pm}S_{a,+,j}|n^{\p}_{r},m^{\p}_{r},
         n^{\p}_{i},m^{\p}_{i}\rangle\end{equation} where
         \begin{equation}\label{fSjjj+1}\begin{array}{l}
         S_{a,+,j}=\sum_{h^{\p}}a_{+,h^{\p}+1,j}
         \ad_{+,h^{\p}+1,j}\ad_{+,h^{\p},j} \\ \hspace{0.3cm}\times
         \sum_{h}a_{+,h+1,j}\ad_{+,h+1,j}a_{+,h,j}a_{+,h,j}
         ,\end{array}\end{equation}
         then Eq. \ref{Nequ} is satisfied. There are three other
         equations one each for $S_{a,-,j},S_{b,+,j}$ and
         $S_{b,-,j}.$

         The expression $=_{\pm}$ denotes equality up to a
         possible sign change.  This can occur because the $S$
         operators are products of an odd number of a-c operators.
         If one wants to implement these state reduction steps
         dynamically with operators that preserve fermion (or
         boson) number, then a pool of additional fermions (or bosons) must be
         available to serve as a source or sink of systems. This
         is not included here because the emphasis is on defining
         complex rational states and  their arithmetic
         properties.

         It is worth noting that Eqs. \ref{jj1}, \ref{jjj+1},
         \ref{fjj1}, and \ref{fjjj+1} can be regarded as axiomatic
         definitions of $=_{N}$ with no reference to their
         numerical meaning in terms of powers of $2$.  The use of
         numbers in their description is included as an aid to the
         reader. It plays no role in their definition.Later
         on a map from the complex states to $C$ will be
         defined that shows that these properties of $=_{N}$ are
         consistent with the map.

         Reduction of a nonstandard representation to a standard
         one proceeds by iteration of steps based on the above
         equivalences.  At some point the process stops when one
         ends up with a state with at most one system of the $a$ or
         $b$ type at each site $j$. This is the case for both
         bosons and fermions.  The possible
         options for each $j$ can be expressed as
         \begin{equation}\label{stdconv}\begin{array}{l}
         |n_{r,j},m_{r,j},n_{i,j},m_{i,j}\rangle
         =\left\{ \begin{array}{l}|1,0,0,1\rangle \\
         |1,0,1,0\rangle\\|0,1,0,1\rangle\\|0,1,1,0\rangle\end{array}
         \mbox { or }\left\{\begin{array}{l}|0,0,0,1\rangle\\
         |0,0,1,0\rangle\\|1,0,0,0\rangle\\|0,1,0,0\rangle\end{array}\right.\right.
        \\ \mbox{} \\ \hspace{3cm} \mbox{ or }|0,0,0,0\rangle.
        \end{array}\end{equation} An example of such a
         state for several $j$ is $|1_{+,3}i_{+,3}1_{-,2}i_{-,4}
         1_{-,-6}\rangle.$ This state corresponds to the number
         $2^{3}-2^{2}-2^{-6}+i(2^{3}-2^{4}).$

         Conversion of a state in this form into a standard state requires
         first determining the signs of the $a$ and $b$ systems occupying the
         sites with the largest $j$ values. This determines the
         signs separately for the real and imaginary components of
         the standard representation.
         In the example given above the real component is $+$ as
         $3>2,-6$ and the imaginary component is $-$ as $4>3.$

         Conversion of all a-c operators into the same kind, as
         shown in Eq. \ref{stdbos}, is based on four relations
         obtained by iteration of Eq. \ref{abjabj} and use of
         Eq. \ref{abdjabj}.  For $k<j$  and for bosons
         they are \begin{equation}\label{abjk}
         \begin{array}{l}a^{\dag}_{+,j}\ad_{-,k} =_{N}a^{\dag}_{+,j-1}\cdots
         a^{\dag}_{+,k} \\ a^{\dag}_{-,j}\ad_{+,k}
         =_{N}a^{\dag}_{-,j-1}\cdots
         a^{\dag}_{-,k} \\ b^{\dag}_{+,j}\bd_{-,k}
         =_{N}b^{\dag}_{+,j-1}\cdots
         b^{\dag}_{+,k} \\ b^{\dag}_{-,j}\bd_{+,k}
         =_{N}b^{\dag}_{-,j-1}\cdots
         b^{\dag}_{-,k}.\end{array}\end{equation}  These equations
         are used to convert all $a$ and all $b$ operators to the
         same type ($+$ or $-$) as the one at the largest occupied
         $j$ value.  Applied to the example $|1_{+,3}i_{+,3}1_{-,2}
         i_{-,4}1_{-,-6}\rangle,$  gives
         $|1_{+,2}1_{+,1}1_{+,0}1_{+,-1}1_{+,-3}\cdots
         1_{+,-6}i_{-,3}\rangle.$ for the standard representation.

         The same four equations hold for fermions provided $h$
         subscripts are included. The values of $h$ are arbitrary
         as they do not affect $=_{N}.$ However, physically,
         application to a state of the form of Eq. \ref{stdconv}
         requires that $h=1$ everywhere, as in $a^{\dag}_{+,1,j}\ad_{-,1,k}
         =_{N}a^{\dag}_{+,1,j-1}\cdots a^{\dag}_{+,1,k}$ for
         example.

         \subsection{Some Useful Operators}\label{SUO}

         Three unitary operators that allow changing between the types of
         systems and moving the string states are useful. For bosons they are
         defined by\begin{equation}\label{WQT} \begin{array}{c}\tilde{W}
         \ad_{+,j}=\ad_{-,j}\tilde{W},\;\;\tilde{W}\bd_{+,j}=\bd_{-,j}\tilde{W}
         \\ \tilde{Q}\ad_{+,j}=\bd_{+,j}\tilde{Q},\;\;\tilde{Q}\ad_{-,j}=
         \bd_{-,j}\tilde{Q} \\ \tilde{T}\cd_{j}=\cd_{j+1}\tilde{T}
         \\ \tilde{W}|0\rangle = |0\rangle,\;\;\tilde{Q}|0\rangle =
         |0\rangle,\;\; \tilde{T}|0\rangle =|0\rangle. \end{array}\end{equation}
         $\tilde{W}$ interchanges $+$ and $-$ states in $r$ and $i$ systems,
         and $\tilde{Q}$ converts $r$ systems to $i$ systems and
         conversely.  $\tilde{T}$ is a translation operator that
         shifts $\ad_{+},\ad_{-},\bd_{+},\bd_{-}$ operator
         products one step along the line of $j$ values.

         For fermions the equations become  \begin{equation}\label{fWQT}
         \begin{array}{c}\tilde{W} \ad_{+,h,j}=\ad_{-,h,j}\tilde{W},
         \;\;\tilde{W}\bd_{+,h,j}=\bd_{-,h,j}\tilde{W}
         \\ \tilde{Q}\ad_{+,h,j}=
         \bd_{+,h,j}\tilde{Q},\;\;\tilde{Q}\ad_{-,h,j}=\bd_{-,h,j}\tilde{Q}
         \\ \tilde{T}\cd_{h,j}=\cd_{h,j+1}\tilde{T}\\ \tilde{W}|0\rangle =
         |0\rangle,\;\;\tilde{Q}|0\rangle =|0\rangle,\;\;\tilde{T}|0\rangle
         =|0\rangle.\end{array}\end{equation}  Note that
         $\tilde{W},\tilde{Q},$ and $\tilde{T}$ commute with one
         another for both bosons and fermions.

         It is  useful to define an operator $\tilde{N}$ that assigns to
         each complex rational state a corresponding
         complex rational number in $C$. For fermions  $\tilde{N}$
         can defined explicitly using a-c operators. One has
         \begin{equation}\label{defN}\begin{array}{l}
         \tilde{N}=\sum_{h,j}2^{j}[\ad_{+,h,j}a_{+,h,j}-
         \ad_{-,h,j}a_{-,h,j} \\ \hspace{1cm}+i(\bd_{+,h,j}
         b_{+,h,j}-\bd_{-,h,j}b_{-,h,j})].\end{array}\end{equation}
         From this definition one can obtain the
         following properties:\begin{equation}\label{Ndef}\begin{array}{c}
         \tilde{N}\tilde{W}+\tilde{W}\tilde{N}=0 \\
         \mbox{$[\tilde{N},\ad_{\a,h,j}]$}
         =\a 2^{j}\ad_{\a,h,j} ;\;\;\;\;\;
         [\tilde{N},\bd_{\b,h,j}]
         =i\b 2^{j}\bd_{\b,h,j}\\ \tilde{N}|0\rangle
         =0.\end{array}\end{equation} Here $\a = +,-$ and $\b =+,-.$
         These equations also apply to bosons if the $h$ variable is
         deleted.

         The function of the operator $\tilde{N}$ is to provide a
         link of complex rational states to the complex
         numbers in $C.$  For each of these states, the $\tilde{N}$ eigenvalue
         is the complex number equivalent, in $C,$ of the complex rational
         number that $\tilde{N}$ associates to these states.

         The eigenvalues of $\tilde{N}$ acting on  states that are
         products of $\ad$ and $\bd$ operators can be obtained from Eqs.
         \ref{defN} or \ref{Ndef}. As an
         example, for the state $\ad_{+,k_{1}}\ad_{-,k_{2}}\bd_{+,k_{3}}\bd_{-,k_{4}}
         |0\rangle,$
         \begin{equation}\label{Nex}\begin{array}{l}
         \tilde{N}\ad_{+,k_{1}}\ad_{-,k_{2}}\bd_{+,k_{3}}\bd_{-,k_{4}}|0\rangle
         = \\ \hspace{0.5cm}(2^{k_{1}}-2^{k_{2}}+i2^{k_{3}}-i2^{k_{4}}) \\
         \hspace{1cm}\times\ad_{+,k_{1}}\ad_{-,k_{2}}\bd_{+,k_{3}}
         \bd_{-,k_{4}}|0\rangle.\end{array}\end{equation}
         For standard representations in general one has
         \begin{equation}\label{NNcs}\tilde{N}(\ad_{\alpha})^{s}(\bd_{\beta})^{t}
         |0\rangle =N[(\ad_{\alpha})^{s}(\bd_{\beta})^{t}](\ad_{\alpha})^{s}
         (\bd_{\beta})^{t}|0\rangle\end{equation} where \begin{equation}\label{Ncs}
         \tilde{N}[(\ad_{\alpha})^{s}(\bd_{\beta})^{t}]=\left\{\begin{array}
         {ll}2^{s}+i2^{t} & \mbox{ if } \alpha=+,\beta=+ \\ -2^{s}+i2^{t} &
         \mbox{ if } \alpha=-,\beta=+ \\ 2^{s}-i2^{t} & \mbox{ if } \alpha=+,\beta=-
         \\ -2^{s}-i2^{t} & \mbox{ if } \alpha=-,\beta=-.
         \end{array}\right.\end{equation} Here $2^{s}=\sum_{j\epsilon
         s}2^{j}$ and $2^{t}=\sum_{k\epsilon t}2^{j}$.

         These results also hold for fermion states. For standard
         states Eq. \ref{stdfer} gives an explicit representation
         for $(\ad_{\alpha})^{s}(\bd_{\beta})^{t}(\ad_{\alpha})^{s}
         (\bd_{\beta})^{t}|0\rangle.$

         The operator $\tilde{N}$ has the satisfying property that any
         two states that  are $N$ equal have the same  $\tilde{N}$
         eigenvalue. If the state $|n_{r},m_{r},n_{i},m_{i}\rangle=_{N}
         (\ad_{\alpha})^{s}(\bd_{\beta})^{t} |0\rangle$ then \begin{equation}
         \tilde{N}|n_{r},m_{r},n_{i},m_{i}\rangle =_{N}
         \tilde{N} (\ad_{\alpha})^{s}(\bd_{\beta})^{t}|0\rangle.\end{equation}
         Here $\alpha =+,-$ and $\beta=+,-.$ This follows from Eqs.
         \ref{abdjabj},\ref{abjabj}, and \ref{defN}.

         These results show that the eigenspaces of $\tilde{N}$
         are invariant for any process of reducing a
         nonstandard state to a standard state using Eqs.
         \ref{abdjabj}-\ref{fabjabj}. Any state
         $|n_{r}m_{r}n_{i}m_{i}\rangle$ with $n_{r,j}\geq 1$ and
         $m_{r,j}\geq 1$ for some $j$ has the same $\tilde{N}$ eigenvalue as the
         state with both $n_{r,j}$ and $m_{r,j}$ replaced by
         $n_{r,j}-1$ and $m_{r,j}-1.$ Also if $n_{r,j}\geq 2$ then
         replacing $n_{r,j}$ by $n_{r,j}-2$ and $n_{r,j+1}$ by
         $n_{r,j+1}+1$ does not change the $\tilde{N}$ eigenvalue.
         Similar relations hold for $m_{r,j},n_{i,j},m_{i,j}.$
         These results show that each eigenspace of $\tilde{N}$
         is infinite dimensional. It is spanned by an infinite
         number of nonstandard complex rational  states
         and exactly one standard state.

         One may think that, because of the association of one
         standard state to each eigenspace, one could
         limit consideration to standard states only.
         As is seen below, this is not the case as  basic arithmetic operations
         generate nonstandard states even when implemented on standard states.

         \section{Basic Arithmetic Operations on Rational
         States}\label{BAORSS}
          \subsection{Addition and Subtraction}\label{AS}

          In quantum mechanics $n-ary$ operations for $n\geq 2$
          are usually represented as operators acting on $n$ fold
          tensor product states of $n$ systems in $n$ different states.
          For many operations $n=2$ or $n=3$ if unitarity is to be
          preserved for the operations. Here, examples include arithmetic
          addition and multiplication where $n=3.$ The operator
          acts on two systems in different states and gives the result of the
          operation as the state of a third system.

          The question arises of how to represent this setup in $\mathcal H^{Ra}$
          using products of boson or fermion a-c operators acting on the vacuum.
          One way is to introduce additional distinguishable
          particles.  For example for fermions, besides the
          operators $\ad_{\a,h,j},\bd_{\b,h,j}$ one has the
          creation operators
          $\hat{\ad}_{\a,h,j},\hat{\bd}_{\b,h,j}$ and
          $\hat{\hat{\ad}}_{\a,h,j},\hat{\hat{\bd}}_{\b,h,j}$ and
          corresponding annihilation operators to
          represent the added fermions. In this case the operators
          for the different types of fermions all commute with one
          another just as the $\ad$ and $\bd$ operators do.

          Another approach is to continue with the two types of
          distinguishable $a$ and $b$
          systems but add additional degrees of freedom to the
          system states. An example would be to consider three
          different regions of space parameterized by an
          additional variable $z.$  In this case arithmetic
          operations would be carried out on systems in $z=1$
          and $z=2$ states and the result given as states of
          systems in $z=3$ states.  For fermions the relevant
          creation operators would be $\ad_{\a,h,j,z},\bd_{\b,h,j,z}$
          with the same type of commutation  relations
          as before (the $\ad s$ and $\bd s$ anticommute among themselves and
          the $\ad s$ commute with the $\bd s$).

          The above approaches also hold for distinguishable and
          indistinguishable bosons except that all the a-c
          operators commute.   In this case the $h$ variable
          is not needed

          In what follows the usual product state representation
          will be used because it is more familiar and is less cumbersome.
          It is left up to the reader to convert
          the states to an a-c operator representation based on
          the above or any other choice of distinguishable and
          indistinguishable systems.

         Addition and multiplication operators that are unitary
         can be defined for complex rational states. For
         addition one has
         \begin{equation}\label{defadd}\begin{array}{l}
         \tilde{+}|n_{r},m_{r},n_{i},m_{i}\rangle|n^{\p}_{r},m^{\p}_{r},
         n^{\p}_{i},m^{\p}_{i}\rangle|0\rangle = \\
         |n_{r},m_{r},n_{i},m_{i}
         \rangle|n^{\p}_{r},m^{\p}_{r},n^{\p}_{i},m^{\p}_{i}\rangle
         |n+n^{\p},m+m^{\p}\rangle.\end{array}\end{equation} Here
         $|n+n^{\p},m+m^{\p}\rangle$ denotes
         $|n_{r}+n^{\p}_{r},m_{r}+m^{\p}_{r},n_{i}+n^{\p}_{i},
         m_{i}+m^{\p}_{i}\rangle$ where the functions
         $n_{r}+n^{\p}_{r}$ correspond to addition of $n_{r}$ and
         $n^{\p}_{r}:$ \begin{equation}\label{compaddn}
         (n_{r}+n^{\p}_{r})_{j}=
         \left\{\begin{array}{l} n_{r,j} \mbox{ if $j$ is in $s$
         and not in $s^{\p}$} \\ n^{\p}_{r,j} \mbox{ if $j$ is not
         in $s$ and is in $s^{\p}$} \\ n_{r,j}+n^{\p}_{r,j} \mbox
         {if $j$ is in $s$ and in $s^{\p}$} \\ 0 \mbox{ otherwise.}
         \end{array}\right. \end{equation} where $s$ and
         $s^{\p}$ are the domains of $n_{r}$ and $n^{\p}_{r}.$
         Similar expressions hold for $(m_{r}+m^{\p}_{r})_{j}
         ,(n_{i}+n^{\p}_{i})_{j},(m_{i}+m^{\p}_{i})_{j}.$

         For bosons or fermions $|n_{r}+n^{\p}_{r},m_{r}+m^{\p}_{r},n_{i}+n^{\p}_{i},
         m_{i}+m^{\p}_{i}\rangle$ is given by Eqs. \ref{occno} and
         \ref{occnost} or \ref{occferm} and \ref{occnoferm} with
         $n_{r,j}$ replaced by $n_{r,j}+n^{\p}_{r,j},$ etc.. Also
         the $j$ product is over all $j$ in the union of the $8$ sets
         $s_{r},s_{i}, s^{\p}_{r},\cdots$ which are the nonzero
         domains of the respective $n$ and $m$ functions.

         For fermions there may be a sign change in the above in case the
         total number of systems in the states $|n_{r},m_{r},n_{i},
         m_{i}\rangle$ and $|n^{\p}_{r},m^{\p}_{r},
         n^{\p}_{i},m^{\p}_{i}\rangle$ is odd. This occurs because
         in this case an odd number of additional systems is
         created by the addition operation.  As noted before, if
         fermion number is to be preserved by dynamical operations,
         then an additional supply needs to be available to serve
         as a source or sink of fermions.

         For standard representations the compact notation
         $|\alpha s, \beta t\rangle = (\ad_{\alpha})^{s}
         (\bd_{\beta})^{t}|0\rangle$ is useful where
         $\alpha =+,-$ and $\beta =+,-.$ One has from Eq.
         \ref{defadd}\begin{equation}\label{defplus}\begin{array}{l}
         \tilde{+}|\alpha s,\beta t\rangle
         |\alpha^{\p}s^{\p},\beta^{\p}
         t^{\p}\rangle|0\rangle= \\ \hspace{1cm}|\alpha s,\beta t\rangle
         |\alpha^{\p}s^{\p},\beta^{\p} t^{\p}\rangle|\alpha s,
         \beta t+\alpha^{\p} s^{\p},\beta^{\p}t^{\p}\rangle\end{array}\end{equation}
         where \begin{equation}\label{alpbetab}\begin{array}{l}|\alpha s,
         \beta t+\alpha^{\p} s^{\p},\beta^{\p}t^{\p}\rangle \\ \hspace{0.5cm}=(\ad_{\alpha})^{s}
         (\bd_{\beta})^{t}(\ad_{\alpha^{\p}})^{s^{\p}}
         (\bd_{\beta^{\p}})^{t^{\p}}|0\rangle =\\ \hspace{1cm}=(\ad_{\alpha})^{s}
         (\ad_{\alpha^{\p}})^{s^{\p}}(\bd_{\beta})^{t}(\bd_{\beta^{\p}})^{t^{\p}}
         |0\rangle \\ \hspace{1.5cm}=|\alpha s+\alpha^{\p} s^{\p},\beta
         t+\beta^{\p}t^{\p}\rangle .\end{array}\end{equation}
         This result, which uses the commutativity of the $a$ and $b$
         a-c operators, shows the separate addition of the $a$
         and  $b$ components of the states.

         It is evident from this that the result of addition need
         not be a standard representation even for standard input
         states. A nonstandard result occurs if  $s$ and $s^{\p},$
         or $t$ and $t^{\p}$ contain one or more elements in common,
         or $\a\neq\ap$ or $\b\neq\bp.$ For
         these cases the methods described would be used to
         reduce the final result to a standard representation.
         Here the reduction is fairly simple as there is  at
         most one application of Eqs. \ref{abdjabj} and \ref{abjabj}
         for each $j$ value. More reduction steps are needed for
         the results of iterated additions.

         From now on the above notation will be used for both
         fermions and bosons with the understanding that for
         fermions the real component $(\ad_{\a})^{s}(\ad_{\ap})^{\sp}$
         is given by Eqs. \ref{occnoferm} and \ref{stdfer} with
         Eq. \ref{compaddn} applying if $\a =\ap.$ Recall that
         for standard representations the functions $n_{r},n_{i},m_{r},m_{i}$
         all have the constant value $1$ over their nonzero
         domains. Also for fermions the equality sign in Eq.
         \ref{defplus} is replaced by $=_{\pm}$ or equality up to
         the sign. If the number of fermions in $|\asbt
         +\asbtp\rangle$ is odd the sign is minus.  Otherwise it
         is even. The sign is always $+$ if the dynamical steps
         of addition conserve the fermion number by use of a
         sink or source of fermions. Also,
         for fermions, the right hand operator products
         $(\ad_{\a})^{s}(\ad_{\ap})^{\sp}(\bd_{\b})^{t}(\bd_{\bp})^{\tp}$
         must  be expressed in the standard order with $j$
         increasing to the right and the appropriate values
         of $n_{r,j}=2$ or $m_{r,j}=2$ in case $\a=\ap$ and $s$ and
         $sp$ have elements in common. A similar situation holds
         for the $\bd$ operator products.

          Extension of $\tilde{+}$  to act on states that are
          linear superpositions of rational string states generates
          entanglement. The discussion will be limited to standard
          states, but it also applies to linear superpositions over
          all states, both standard and nonstandard.
          Let $\psi=\sum_{\alpha,
         s,\beta,t}d_{\alpha,s,\beta,t}|\alpha s,\beta t\rangle$
         and $\psi^{\p}=\sum_{\alpha^{\p},
         s^{\p},\beta^{\p},t^{\p}}d^{\p}_{\alpha^{\p},s^{\p},
         \beta^{\p},t^{\p}}|\alpha^{\p} s^{\p},\beta^{\p}
         t^{\p}\rangle.$ Then
          \begin{equation}\label{plusentngl}\begin{array}{l}
          \tilde{+}\psi\,\psi^{\prime}|0\rangle =\sum_{\alpha
          ,s,\beta,t}\sum_{\alpha^{\p},s^{\p},\beta^{\p},t^{\p}}
          d_{\alpha, s,\beta,t}d^{\p}_{\alpha^{\p},
          s^{\p},\beta^{\p},t^{\p}} \\ \hspace{1cm}\times|\alpha s,\beta t\rangle
          |\alpha^{\p}s^{\p},\beta^{\p}t^{\p}\rangle|\alpha s,\beta t +\alpha^{\p}
          s^{\p},\beta^{\p}t^{\p}\rangle\end{array} \end{equation} which is entangled.

          To describe repeated arithmetic operations
          it is useful to have a state that describes directly the
          addition of $\psi$ to $\psi^{\prime}$. Since the overall
          state shown in Eq. \ref{plusentngl} is entangled,
          the desired state would be expected to be a mixed or
          density operator state.  This is indeed the case as can
          be seen by taking the trace over the first two
          components of $\tilde{+}\psi,\psi^{\p}|0\rangle:$
          \begin{equation}\label{rhoaddcmplx}\begin{array}{l}
          \rho_{\psi+\psi^{\prime}}=Tr_{1,2}\tilde{+}
          |\psi\rangle|\psi^{\p}\rangle|0\rangle\langle 0|\langle\psi^{\p}
          |\langle\psi|\tilde{+}^{\dag}= \sum_{\alpha,\beta, s,t}
          \\ \times\sum_{\alpha^{\p},\beta^{\p},s^{\p},t^{\p}}
          |d_{\alpha, s,\beta,t}|^{2}|d^{\p}_{\alpha^{\p},s^{\p},\beta^{\p},t^{\p}}|^{2}
          \rho_{\alpha s,\beta t+ \alpha^{\p}s^{\p},\beta^{\p}t^{\p}}.\end{array}
          \end{equation} Here $\rho_{\alpha s,\beta t+
          \alpha^{\p}s^{\p},\beta^{\p}t^{\p}}$ is the pure state
          density operator $|\alpha s,\beta
          t+ \alpha^{\p}s^{\p},\beta^{\p}t^{\p}\rangle\langle\alpha s,\beta
          t+ \alpha^{\p}s^{\p},\beta^{\p}t^{\p}|.$
          The expectation value of $\tilde{N}$ on this state gives
          the expected result:
          \begin{equation}\label{Nplus} Tr(\tilde{N}\rho_{\psi+\psi^{\prime}})
          =\langle\psi|\tilde{N} |\psi\rangle+\langle\psi^{\prime}
          |\tilde{N}|\psi^{\prime}\rangle.\end{equation}

          For subtraction use is made of the fact that
          $|\alpha^{\p}s,\beta^{\p}t\rangle$ is the additive
          inverse of $|\alpha s,\beta t\rangle$ if $\alpha^{\p}\neq\alpha$
          and $\beta^{\p}\neq\beta.$ Then \begin{equation}\label{addinv}
          |\alpha s,\beta t+\alpha^{\p}s,\beta^{\p}t\rangle =_{N}|0\rangle
          \end{equation} where Eq. \ref{abdjabj} is used to give $(\ad_{\alpha})^{s}
          (\ad_{\alpha^{\p}})^{s}=_{N}1=_{N}(\bd_{\beta})^{t}
          (\bd_{\beta^{\p}})^{t}.$

          A unitary subtraction operator, $\tilde{-},$
          is defined by \begin{equation}\label{subtr}\begin{array}{l}
          \tilde{-}|\alpha s,\beta t\rangle |\alpha^{\p}s^{\p},\beta^{\p}
         t^{\p}\rangle|0\rangle= \\ \hspace{0.5cm}|\alpha s,\beta t\rangle
         |\alpha^{\p}s^{\p},\beta^{\p} t^{\p}\rangle|\alpha
         s,\beta t-\alpha^{\p} s^{\p},\beta^{\p} t^{\p}\rangle\end{array}\end{equation}
         where $|\alpha s,\beta t-\alpha^{\p} s^{\p},\beta^{\p} t^{\p}\rangle
         = |\alpha s,\beta t+\alpha^{\p\p} s^{\p},\beta^{\p\p}
         t^{\p}\rangle$ and $\alpha^{\p\p}\neq\alpha^{\p}$ and
         $\beta^{\p\p}\neq \beta^{\p}.$ Other properties of $\tilde{-},$
         including extension to nonstandard states and linear state
         superposition, are similar to those for addition.

          One sees that the definition of $\tilde{+}$,
          Eqs. \ref{defadd}-\ref{defplus}, satisfies the requisite
          properties of addition. it is commutative
          \begin{equation}\begin{array}{l}(a^{\dag}_{\alpha})^{s}(\bd_{\beta})^{t}
          (\ad_{\alpha^{\p}})^{s^{\prime}}(\bd_{\beta^{\p}})^{t^{\p}}|0\rangle=_{N}
          \\ \hspace{1cm}(\ad_{\alpha^{\p}})^{s^{\prime}}(\bd_{\beta^{\p}})^{t^{\p}}
          (a^{\dag}_{\alpha})^{s}(\bd_{\beta})^{t}|0\rangle\end{array}\end{equation}
          and associative\begin{equation}\begin{array}{l}
          (\ad_{\alpha})^{s}(\bd_{\beta})^{t}
          \{(a^{\dag}_{\alpha^{\p}})^{s^{\prime}}
          (\bd_{\beta^{\p}})^{t^{\p}}(\ad_{\alpha^{\p\p}})^{s^{\p\p}}
          (\bd_{\beta^{\p\p}})^{t^{\p\p}}\}|0\rangle  \\ =_{N}
          \{ (\ad_{\alpha})^{s} (\bd_{\beta})^{t}(a^{\dag}_{\alpha^{\p}})^{s^{\p}}
          (\bd_{\beta^{\p}})^{t^{\p}}\}(\ad_{\alpha^{\p\p}})^{s^{\p\p}}
          (\bd_{\beta^{\p\p}})^{t^{\p\p}}|0\rangle.\end{array}\end{equation}
          Also $|0\rangle$ is the additive identity.
          This is expressed here by noting that $(\ad_{\a})^{s}=(\bd_{\b})^{t}=1$
          if $s$ or $t$ are empty.  Note that these properties are expressed in terms
          of $N$ equality, not state equality as these properties
          may not hold for state equality. For example,
          for fermions, the minus sign introduced by operator commutation
          has no effect on the numerical value. but it can have a
          nontrivial consequence for linear superposition states.
          However, even in this case it does not affect the
          numerical properties of states such as $\rho_{\psi+\psi^{\p}}.$
          For bosons there is no problem because the a-c operators commute.
          Also the properties of $N$ equality are useful to show
          that associativity, etc., also hold for addition of nonstandard states.

          \subsection{Multiplication}\label{M}

          The description of multiplication is more complex
          because it is an iteration of addition, and
          complex rational states are involved.  The operator
          $\tilde{\times}$ is defined by
          \begin{equation}\label{timescmplx}\begin{array}{l}
          \tilde{\times}|\alpha s, \beta t\rangle|\alpha^{\prime}
          s^{\prime},\beta^{\p} t^{\p}\rangle |0\rangle = \\ \hspace{0.5cm}|\alpha
          s,\beta t\rangle|\alpha^{\prime}s^{\prime},\beta^{\p}t^{\p}\rangle
          |\alpha s, \beta t  \times \alpha^{\prime}s^{\prime},
          \beta^{\p}t^{\p}\rangle.\end{array} \end{equation}

          The definition of the state $|\alpha s,\beta t\times \alpha^{\prime}s^{\prime},
          \beta^{\p}t^{\p}\rangle$ is most easily expressed
          as follows. Let $c_{j}$ stand for any one of
          $\ad_{+,j},\ad_{-,j},\bd_{+,j},\bd_{-,j}$ and $c^{s}$ for any one of
          $(\ad_{+})^{s},(\ad_{-})^{s},(\bd_{+})^{s},
          (\bd_{-})^{s}.$ This use of a variable without a dagger
          to represent any one of the four creation operators is
          done in the following to avoid symbol clutter.

          The definition of multiplication is divided into two
          steps: converting the product $c^{s}\times \hc^{t}$
          into a product of $c_{0}$ times some operator product and
          then defining $c_{0}\times --$ to take account of
          complex numbers. Note that the $\tilde{N}$ eigenvalues of
          $c_{0}|0\rangle$ range over the numbers $1,-1,i,-i.$

          The first step uses Eq. \ref{WQT} to define $c_{j}\times
          \hc^{s}$ by
          \begin{equation}\label{cjmult}\begin{array}{l}
          c_{j}\times \hc^{s}=c_{0}\times
          \tilde{T}^{j}\hc^{s}(\tilde{T}^{\dag})^{j}= \\ \hspace{0.5cm}(c_{0}\times
          \hc_{k_{1}+j})\cdots (c_{0}\times
           \hc_{k_{n}+j}).\end{array}\end{equation}
           Here $s=\{k_{1},k_{2},\cdots,k_{n}\}$ where
           $k_{1}<k_{2}<\cdots<k_{n}$ for fermions. This equation
           shows that multiplication by a power of $2$
          is equivalent to a  $j$ translation by that power.

           Extension of this to multiplication by a product of the
           $c$ operators gives \begin{equation}\label{cscspmult}
          \begin{array}{l}c^{s}\times\hc^{s^{\p}} = (c_{k_{1}}\times
          \hc^{s^{\p}})(c_{k_{2}}\times \hc^{s^{\p}})\cdots (c_{k_{n}}\times
          \hc^{s^{\p}}) \\ =(c_{0}\times\tilde{T}^{k_{1}}\hc^{s^{\p}}
          (\tilde{T}^{\dag})^{k_{1}})\cdots (c_{0}\times\tilde{T}^{k_{n}}\hc^{s^{\p}}
          (\tilde{T}^{\dag})^{k_{n}}).\end{array} \end{equation}Here
          $s=k_{1},k_{2},\cdots ,k_{n}.$

          For the second step, all four cases of multiplication by
          $c_{0}$ can be expressed as:  \begin{equation}\label{ab0mult}
          \begin{array}{l}\bd_{-,0}\times c^{s} =\left\{\begin{array}{ll}\tilde{Q}
          \tilde{W}c^{s} \tilde{W}^{\dag}\tilde{Q}^{\dag} & \mbox{
          if }c=\ad_{+}, \ad_{-} \\ \tilde{Q}c^{s}\tilde{Q}^{\dag} & \mbox{ if
          }c=\bd_{+},\bd_{-}.\end{array}\right. \\
          b^{\dag}_{+,0}\times c^{s} =\left\{\begin{array}{ll}\tilde{Q}
          c^{s}\tilde{Q}^{\dag} & \mbox{ if }c=\ad_{+}, \ad_{-} \\\tilde{Q}
          \tilde{W}c^{s} \tilde{W}^{\dag}\tilde{Q}^{\dag} & \mbox{ if }
           c=\bd_{-},\bd_{+}.\end{array}\right.\end{array} \end{equation} For all $c$
           \begin{equation}\label{a0mult}\begin{array}{l} \ad_{-,0}\times c^{s}=
           \tilde{W}c^{s}\tilde{W}^{\dag} \\
           \ad_{+,0}\times c^{s}=c^{s} \\  c_{0}\times\tilde{1}=
          \tilde{1}\times c_{0}=\tilde{1}. \end{array}\end{equation}
           This gives
          \begin{equation}\label{asxbt}\begin{array}{l}
          |\alpha s, \beta t  \times \alpha^{\prime}s^{\prime},
          \beta^{\p}t^{\p}\rangle =\\ |(\alpha s\times
          \alpha^{\prime}s^{\prime}+\beta t\times
          \beta^{\p}t^{\p}),(\alpha s\times \beta^{\p}t^{\p}+\beta
          t\times\alpha^{\prime}s^{\prime}) \rangle.\end{array}\end{equation}
          where $(\alpha s\times\alpha^{\prime}s^{\prime}+\beta t\times
          \beta^{\p}t^{\p})$ and $ (\alpha s\times \beta^{\p}t^{\p}+\beta
          t\times\alpha^{\prime}s^{\prime})$ denote the real and
          imaginary components of the product state. Note that if $s$
          is empty, then $c^{s}= \tilde{1}.$  From the above and Eq. \ref{ab0mult}
          one has $\tilde{1}\times (c)^{s_{\prime}}|0\rangle =
          (c)^{s_{\prime}}\times\tilde{1}|0\rangle = |0\rangle.$
          This corresponds to a proof for the number
          representation constructed here that multiplication
          of any number by $0$ gives $0$.

         Extension of $\tilde{\times}$ to cover nonstandard states
         is straight forward. To see this one notes that any
         nonstandard state can be written in the form
         $c_{1}^{s_{1}}c_{2}^{s_{2}}\cdots c_{n}^{s_{n}}|0\rangle$
         where for each $\ell=1,2,\cdots,n$ $c^{s_{\ell}}_{\ell}$ is any
         one of $(\ad_{+})^{s_{\ell}},(\ad_{-})^{s_{\ell}},
         (\bd_{+})^{s_{\ell}},(\bd_{-})^{s_{\ell}}$ and $s_{\ell}$ is a finite
         set of integers. The product of this state with another
         nonstandard state $\hc_{1}^{t_{1}}\hc_{2}^{t_{2}}\cdots
         \hc_{m}^{t_{m}}|0\rangle$ is the state
         $$\prod_{j=1}^{n}\prod_{k=1}^{m}c_{j}^{s_{j}}\times
         \hc_{k}^{t_{k}}|0\rangle.$$ Each component $c_{j}^{s_{j}}\times
         \hc_{k}^{t_{k}}$ is evaluated according to the description
         in Eqs. \ref{cscspmult} \emph{et seq}. The large number of
         multiplications needed here suggests that it may be more
         efficient to convert each nonstandard state to a standard
         state and then carry out the multiplication.

          Extension of multiplication to linear superpositions of
          complex rational string states is straightforward.
          Following Eq. \ref{rhoaddcmplx}  the result of multiplying
          $\psi$ and $\psi^{\p}$ is the density operator
          $\rho_{\psi\times\psi^{\prime}}$ where \begin{equation}
         \label{rhomultcmplx}\begin{array}{l}
          \rho_{\psi\times\psi^{\prime}}=Tr_{1,2}\tilde{\times}
          |\psi\rangle|\psi^{\p}\rangle|0\rangle\langle 0|\langle\psi^{\p}
          |\langle\psi|\tilde{\times}^{\dag}= \sum_{\alpha,\beta,
          s,t} \\ \times\sum_{\alpha^{\p},\beta^{\p},s^{\p},t^{\p}}
          |d_{\alpha, s,\beta,t}|^{2}|d^{\p}_{\alpha^{\p},s^{\p},\beta^{\p},t^{\p}}|^{2}
          \tilde{P}_{\alpha s\beta t\times \alpha^{\p}s^{\p}\beta^{\p}t^{\p}}.\end{array}
          \end{equation} This is the same as Eq. \ref{rhoaddcmplx} for
          addition except that the projection operator is for the product state
          $|\alpha s\beta t\times
          \alpha^{\p}s^{\p}\beta^{\p}t^{\p}\rangle$.

          From the definition of $\tilde{N}$ one obtains
          \begin{equation}\label{Nmult}
          Tr\tilde{N}\rho_{\psi\times\psi^{\p}}=\langle\psi|\tilde{N}
          |\psi\rangle\langle\psi^{\p}|\tilde{N}|\psi^{\p}\rangle.
          \end{equation} Here $N(\alpha s\beta t\times \alpha^{\p}
          s^{\p}\beta^{\p}t^{\p})= N(\alpha s\beta t)
          N(\alpha^{\p}s^{\p}\beta^{\p}t^{\p})$ has been used.

          The above results also show that multiplication is commutative in that
          \begin{equation}\label{multcomm}\begin{array}{l}
         (\ad_{\alpha})^{s}(\bd_{\beta})^{t}\times
         (\ad_{\alpha^{\p}})^{s^{\p}}\bd_{\beta^{\p}})^{t^{\p}}|0\rangle
         =_{N}\\ \hspace{1cm}(\ad_{\alpha^{\p}})^{s^{\p}}\bd_{\beta^{\p}})^{t^{\p}}\times
         (\ad_{\alpha})^{s}(\bd_{\beta})^{t}|0\rangle.\end{array}\end{equation}
         Distributivity of multiplication over addition for complex
         rational states follows
         from Eqs. \ref{cjmult} and \ref{cscspmult}. To see this
         let $s^{\p}=s_{1}\bigcup s_{2}$ be a partition of
         $s^{\p}$ into two sets where all integers in $s_{1}$ are
         larger than those in $s_{2}.$  Then the equations show
         that  \begin{equation}\begin{array}{c}c^{s}\times \hc^{s^{\p}}=
         c^{s}\times (\hc^{s_{1}}\hc^{s_{2}})=c^{s}\times(\hc^{s_{1}}+
         \hc^{s_{2}}) \\ =_{N}(c^{s}\times\hc^{s_{1}})(c^{s}\times
         \hc^{s_{2}})=(c^{s}\times\hc^{s_{1}})+c^{s}\times \hc^{s_{2}}).
         \end{array}\end{equation} Note again that $N$ equality is
         used, not state equality.

           \subsection{Division}\label{D}
          As is well known the complex rational  string states and linear
          superpositions of these states are
          not closed under division.  However they just escape
          being closed in that division can be approximated to any
          desired accuracy.  One defines an $\ell$ accurate division operator
          $\tilde{\div}_{\ell}$ by
          \begin{equation}\label{defdiv}\begin{array}{l}
          \tilde{\div}_{\ell}|\alpha s, \beta t\rangle|\alpha^{\prime}
          s^{\prime},\beta^{\p} t^{\p}\rangle |0\rangle = \\ \hspace{0.5cm}|\alpha
          s,\beta t\rangle|\alpha^{\prime}s^{\prime},\beta^{\p}t^{\p}\rangle
          |\alpha s, \beta t /(  \alpha^{\prime}s^{\prime},
          \beta^{\p}t^{\p})_{\ell}\rangle\end{array} \end{equation} where \begin{equation}
          |\alpha s, \beta t /(  \alpha^{\prime}s^{\prime},
          \beta^{\p}t^{\p})_{\ell}\rangle =|\alpha s, \beta t\times
          (\alpha^{\prime}s^{\prime},\beta^{\p}t^{\p})^{-1}_{\ell}\rangle\end{equation}
          and \begin{equation}|(\alpha^{\prime}s^{\prime},
          \beta^{\p}t^{\p})^{-1}_{\ell}\rangle=|(\alpha^{\prime}s^{\prime},
          \beta^{\p\p}t^{\p}\times(q^{-1})_{\ell}\rangle.\end{equation} Here
          $\beta^{\p\p}\neq \beta^{\p}$ and
          $q=((\alpha^{\p}s^{\p})^{2}+\beta^{\p}t^{\p}\times
          \beta^{\p\p}t^{\p})^{1/2}_{\ell}.$ This is the complex
          rational state expression of
          $(u+iv)^{-1}=(u-iv)/(u^{2}+v^{2})^{1/2}.$

          Determination of the real rational state
          $|q^{-1}_{\ell}\rangle$ involves two computations to
          accuracy $\ell,$ a square root and an inverse. Since
          "accuracy $\ell$" is common to both, The discussion here
          will be limited to the inverse as the square root is
          calculated in a similar way but with a different
          algorithm.

          The above shows it is sufficient to consider
          states of the form $(\ad_{+})^{s}|0\rangle$ or $(\ad_{-})^{s}|0\rangle$
          in detail as extension to imaginary and complex rational
          states uses these results. The main goal is to show that
          any string state $(\ad_{+})^{s}|0\rangle$ or $(\ad_{-})^{s}|0\rangle$ has
          an inverse string state to an accuracy of at least
          $\ad_{+,-\ell}|0\rangle$ for any $\ell.$ Accuracy is defined by means
          of an ordering relation $<_{N}$ on the rational string number states.
          A few details are given in the next section. The inverse can be used with
          the definition of multiplication to show
          that\begin{equation}\label{cdivcp}(\cs/\hc^{s^{\p}})_{\ell}
          |0\rangle=(\cs\times(\hc^{s^{\p}})^{-1}_{\ell})|0\rangle
          \end{equation} where $c,\hc$ are each  either
          $\ad_{+}$ or $\ad_{-}.$ This would be applied to
          $[(\alpha^{\p}s^{\p},\beta^{\p\p}t^{\p}))/q]_{\ell}$
          to evaluate $|(\alpha s,\beta t)/(\alpha^{\p}s^{\p},
          \beta^{\p}t^{\p})\rangle.$

          Let $\ad_{+,[-1,-\ell]}=\ad_{+,-1}\ad_{+,-2}\cdots
          \ad_{+,-\ell}.$  One has to show that for each operator
          product $(\ad_{\alpha})^{s}$ and each $\ell$ there exists a
          product $(\ad_{\alpha})^{t},$ where \begin{equation}\label{definv}
          (\ad_{\alpha})^{s}\times(\ad_{\alpha})^{t}|0\rangle =_{N}
          \ad_{+,[-1,-\ell]}\ad_{+,<-\ell}|0\rangle.\end{equation}
          Here $\ad_{+,<-\ell}$ is
          an arbitrary product of $\ad_{+}$ operators at locations $<-\ell.$
          The arithmetic difference between the states $\ad_{+,[-1,-\ell]}
          \ad_{+,<-\ell}|0\rangle$
          and $\ad_{+,0}|0\rangle$ is less than $\ad_{+,-\ell}|0\rangle.$

           For each $(\ad_{\alpha})^{s}$ and each $\ell$, the inverse
           product $(\ad_{\alpha})^{t}$ can be constructed inductively.
          Details are given in the appendix. Extension of the
          definition to cover division to accuracy $\ell$ by a
          nonstandard state may be possible in principle but an
          inductive construction of the inverse of a nonstandard
          state seems prohibitive.  In this case it is much more
          efficient to convert the nonstandard state to a standard
          one and then construct the inverse following the methods
          in the appendix. The result obtained will be $N$ equal
          to the direct inverse of the nonstandard state.

          As was done  for addition and multiplication, a unitary
          operator for division to accuracy $\ell$ on linear superposition
          states can be defined. The result, $\rho_{\psi/\psi^{\p}},$
          given by Eq. \ref{rhomultcmplx} with
          $\tilde{P}_{(\alpha s\beta t)/
          (\alpha^{\p}s^{\p}\beta^{\p}t^{\p})_{\ell}}$ replacing
          $\tilde{P}_{\alpha s\beta t\times
          \alpha^{\p}s^{\p}\beta^{\p}t^{\p}},$ is obtained by
          tracing over the fist two states.

          It is to be noted that arithmetic operations on complex
          rational states satisfy the necessary properties, such as commutativity,
          distributivity, existence of an $\ell$ inverse, etc. However
          linear superposition states do not satisfy all these properties.
          No triple $\psi,\psi^{\p},\psi^{\p\p},$ satisfies the distributive law
         \begin{equation}\label{supdist}\psi\times_{N}\psi^{\p}+_{N}\psi
         \times_{N}\psi^{\p\p}=_{N}\psi\times_{N}(\psi^{\p}+_{N}\psi^{\p\p}).
         \end{equation} Also linear superposition states do
          not have $\ell$ inverses. Given $\psi$ there is no state
          $\psi^{\p}_{\ell}$ that satisfies\begin{equation}\label{supinv}
          \psi\times\psi^{\p}=\ad_{+,[-1,-\ell]}\ad_{+,<-\ell}|0\rangle.
          \end{equation}

          \section{Discussion}

          In this paper a binary quantum mechanical representation of complex
          rational  numbers was presented that did not use
          qubits. It is based on the observation that the
          numerical value of a qubit state such as
          $|10010.01\rangle$ depends on the distribution of $1s$
          only with the $0s$ functioning merely as place holders.

          The representation described here extends the literature
          representations \cite{BenRNQM,BenRNQMALG,Kitaev} to include
          boson and fermion representations of complex rational
          numbers.  The representation is compact and seems well
          suited to represent complex
          rational numbers. Since both standard and nonstandard representations
          are included, arithmetic combinations of different types of numbers
          are relatively easy to represent.  This is not the case for
          qubit product states, which are limited to standard
          representations. For example the qubit
          representation of the nonstandard state
          $\ad_{-,-1}\bd_{-,6}\bd_{+,2}\ad_{+,3}|0\rangle$
          is the pair of states, $|111.1\rangle,|-i111100.0\rangle.$
          These qubit states correspond to the standard representations,
          $\ad_{+,2}\ad_{+,1}\ad_{+,0}\ad_{+,-1}|0\rangle$ and
          $\bd_{-,5}\bd_{-,4}\bd_{-,3}\bd_{-,2}|0\rangle$ of
          $\ad_{-,-1}\ad_{+,3}|0\rangle$ and $\bd_{-,6}\bd_{+,2}|0\rangle$.

          This flexibility makes the arithmetic operations
          relatively easy to express in that the various steps can
          be shown in a compact form.  For instance addition of
          several complex rational states consists of converting
          a product of creation operator products,  to a standard
          form. This conversion process is equivalent to the steps
          one goes through in carrying out the addition of several
          product qubit states where each product state can be any
          one of the four types of numbers.

          Another advantage for the number representation shown here
          is that it may expand the search horizon for
          implementable physical models of quantum computers.  An
          example of such a model using two types of bosons
          that have two different internal states, $+,-$,  consists of a
          string of Bose Einstein condensate (BEC) pools along an
          integer $j$ lattice. Each  pool can contain up to four
          different BECs where the pool at site $j$ contains $n_{+,j}$
          and $n_{-,j}$ bosons of type $r$ and $m_{+,j}$ and $m_{-,j}$
          bosons of type $i.$

          Such a string of BEC pools is a possible physical model of a
          nonstandard complex rational state. For example, one might
          imagine starting out a quantum computation of $\int_{a}^{b}f(x)dx$
          with all pools empty,  coherently computing many values of
          $f(x_{j})$ for $j=1,\cdots,M,$ and putting the results into the
          pools by adding bosons of the appropriate type and state at
          specified $j$ locations. The resulting string of BEC
          pools is a nonstandard representation of the value of
          the integral.  It is converted to a standard
          representation by removing bosons according to rules
          based on Eqs. \ref{abdjabj} and \ref{abjabj}. This
          corresponds to carrying out the sum indicated by the integral.
          It would be very useful for this conversion if bosons could be
          found that interact physically according to one or more of these rules.
          Then part of the conversion process could happen
          automatically.

          It should be emphasized that the operator $\tilde{N}$
          was introduced early in the development as an aid to
          understanding. It is not essential in that the whole
          development here can be carried out with no reference to
          $\tilde{N}.$ The advantage of this is that complex
          rational states and the arithmetic operations
          can be defined  independently of and without
          reference to corresponding properties on $C$.

          In this case Eqs. \ref{abdjabj} and
          \ref{abjabj}, or Eqs. \ref{fabdjabj} and \ref{fabjabj},
          become definitions of $=_{N}.$  Also the
          definitions of operators for basic arithmetic operations,
          Eqs. \ref{defplus}, \ref{subtr}, \ref{timescmplx}, and
          \ref{defdiv} do not depend on $\tilde{N}.$  As an
          operator on the states in $\mathcal H^{Ra},$
          $\tilde{N}$ corresponds to a map from states $\psi$
          where the expectation value $\langle
          \psi|\tilde{N}|\psi\rangle$ is the number in $C$
          associated to $\psi.$  Eqs. \ref{Nplus} and \ref{Nmult},
          give the satisfactory result that
          $\tilde{N}$ is a morphism from  states in $\mathcal H^{Ra}$ to $C$
          in that it preserves the basic arithmetic operations.

          If $\tilde{N}$ is not used, one needs to define an ordering $<_{N}$
          that satisfies ordering axioms for rational numbers separately on the
          real and imaginary parts. The ordering is  defined on the standard
          positive rational states and extended to the standard negative
          states by reflection. Extension to nonstandard  states uses $=_{N}$
          as in \begin{equation}\begin{array}{l} \mbox{If $|n_{+},n_{-},m_{+},m_{-}
          \rangle =_{N}|\alpha s,\beta t\rangle,$}\\ \hspace{0.25cm}\mbox{
          $|n^{\p}_{+},n^{\p}_{-},m^{\p}_{+},m^{\p}_{-}\rangle
          =_{N}|\alpha^{\p} s^{\p},\beta^{\p} t^{\p}\rangle,$} \\ \hspace{0.5cm}
          \mbox{ and $|\alpha s,\beta t\rangle<_{N}|\alpha^{\p}
          s^{\p},\beta^{\p} t^{\p}\rangle,$ } \\ \hspace{1cm}
          \mbox{ then $|n_{+},n_{-},m_{+},m_{-} \rangle <_{N}|n^{\p}_{+},
          n^{\p}_{-},m^{\p}_{+},m^{\p}_{-}\rangle.$}\end{array}\end{equation}
          The description of division  used $<_{N}$
          implicitly in referring to division to accuracy
          $\ad_{-\ell}|0\rangle$ instead of accuracy $2^{-\ell}.$

          The description given here is not limited
          to binary representations. For a $k-ary$ representation
          one replaces Eq. \ref{abjabj} by \begin{equation}\label{kary}
          (\cdj)^{k}=_{N} \cd_{j+1};\;\;\;\; (\cj)^{k}=_{N}c_{j+1}
          \end{equation} and changes Eq. \ref{abjk} to reflect this
          difference.  Appropriate changes would be needed in any
          results depending on these equations.

          \section*{Acknowledgements}
          This work was supported by the U.S. Department of Energy,
          Office of Nuclear Physics, under Contract No. W-31-109-ENG-38.

            \begin{appendix}
            \section{Appendix}
            One method of constructing a string $(\ad_{+})^{t}$
            inverse to $(\ad_{+})^{s}$ follows a basic procedure.
            The procedure applies also to $(\ad_{-})^{s}$ if $-$ replaces $+$ everywhere in the
            following. In what follows the vacuum state $|0\rangle$ is
            suppressed, and $+$ and $\times$ denote arithmetic operations.

            Let $(\ad_{+})^{s}=\ad_{+,k_{1}},\ad_{+,k_{2}},\cdots,\ad_{+,k_{n}}$ where
          $k_{1}>k_{2}>,\cdots ,>k_{n}.$  The string $(\ad_{+})^{t}=\ad_{+,h_{1}}
          \ad_{+,h_{2}},\cdots$ where $h_{1}>h_{2}>h_{3}\cdots$
          is constructed using the definition of multiplication.
          The first location $h_{1}$ is given by
          $h_{1}+k_{1}=-1.$ For the next location, let $m^{\p}$ be the least $m$ where a
          gap occurs in the descending $k$ values, i.e. the least
          $m$ where $k_{m}<k_{m-1}-1.$ $h_{2}$ is defined by
          $h_{2}+k_{1}=-m^{\p}.$ After converting the string $(\ad_{+})^{s}\times
          \ad_{+,h_{1}}\ad_{+,h_{2}}$ to a standard representation, which
          should contain the initial segment $\ad_{+,[-1,-m]},$ one
          repeats the construction for the next location gap at $m^{\p\p}$
          in the product string. The choice of  $h_{3}$ is determined by
          $h_{3}+k_{1}=-m^{\p\p}$ provided that multiplication of the
          product string $(\ad_{+})^{s}\times \ad_{+,h_{1}}\ad_{+,h_{2}}$ by
          $\ad_{+,h_{3}}$ and conversion to a standard representation
          does not cause the product to become $>\ad_{+,0}.$ If
          this happens then one decreases $h_{3}$ one unit at a
          time until a satisfactory value is found.

          The process is iterated until the resulting standard product
          has the form $\ad_{+,[-1,-\ell]}\ad_{+,<-\ell}.$ Then the process stops.
          The string inverse to $(\ad_{+})^{s}$
          is given by $\ad_{+,h_{1}}\ad_{+,h_{2}}\cdots \ad_{+,h_{n(\ell)}}$
          where $n(\ell)$  is the smallest number such that
         \begin{equation} \begin{array}{l}(\ad_{+})^{s}\times \ad_{+,h_{1}}\ad_{+,h_{2}}\cdots
          \ad_{+,h_{n(\ell)}}=_{N} \\ \hspace{1cm}\ad_{+,[-1,-\ell]}\ad_{+,<-\ell}.
          \end{array}\end{equation}

           As an  example,  let $(\ad_{+})^{s}=
          \ad_{+,-4}\ad_{+,-6}\ad_{+,-9}$ and $\ell=5.$ From the above
          and $m^{\p}=2,$  $h_{1}=3$ and $h_{2}=2.$
          The standard result is
          $\ad_{+,-1}\ad_{+,-2}\ad_{+,-3}\ad_{+,-4}\ad_{+,-6}\ad_{+,-7}.$ Here
          $m^{\p\p}=5$ so one might set $h_{3}+k_{1}=-5$ or
          $h_{3}=-1.$  This will not work because multiplication gives
          \begin{equation}\begin{array}{l}\ad_{+,-1}\ad_{+,-2}\ad_{+,-3}\ad_{+,-4}
          \ad_{+,-6}\ad_{+,-7}+ \\ \hspace{0.5cm}(\ad_{+,-4}
          \ad_{+,-6}\ad_{+,-9})\times \ad_{+,-1} \\ \hspace{1cm}=_{N}  \ad_{+[-1,-6]}
          \ad_{+,-7}\ad_{+,-7}\ad_{+,-10}\end{array}\end{equation}
          and conversion  to a standard form gives a
          string $>_{N}\ad_{+,0}.$ A stepwise decrease in $h_{3}$ gives  $h_{3}=-2$
           which works. The process terminates because the product is
          $\ad_{+,[-1,-5]}\ad_{+,-7}\ad_{+,-8}\ad_{+,-11}.$

            \end{appendix}

            \end{document}